# Electrically Pumped Terahertz Frequency Comb Based on Actively Mode-locked Resonant Tunneling Diode

Feifan Han[1]†, Xiongbin Yu[1]†*, Qun Zhang[1], Zebin Huang[1], Longhao Zou[1], Weichao Li[1], Jingpu Duan[1], Zhen Gao[3]*, and Xiaofeng Tao[2,1]*

[1]Department of Broadband Communication, Pengcheng Laboratory, Shenzhen 518057, China.

[2]National Engineering Research Center of Mobile Network Technologies, Beijing University of Posts and Telecommunications, Beijing 100876, China.

[3]State Key Laboratory of Optical Fiber and Cable Manufacture Technology, Department of Electronic and Electrical Engineering, Guangdong Key Laboratory of Integrated Optoelectronics Intellisense, Southern University of Science and Technology, Shenzhen 518055, China.

†These authors contributed equally to this work.

*Corresponding author. Email: yuxb@pcl.ac.cn (X.B.Y.); gaoz@sustech.edu.cn (Z.G.); taoxf@bupt.edu.cn (X.F.T.)

**Terahertz (THz) frequency combs (TFCs) are promising for numerous applications in spectroscopy, metrology, sensing, and wireless communications. However, the practical applications of TFCs have been hindered thus far by the need for cryogenic cooling, limited bandwidth, and bulky configuration, largely due to the lack of advanced THz sources. Here, we report an electrically pumped, room-temperature, broadband, compact, and tunable TFC by integrating an actively mode-locked resonant tunneling diode (RTD) in a metallic waveguide. By injecting a strong, continuous-wave radio frequency (RF) signal into a direct-current-driven RTD oscillator, we experimentally demonstrate a broadband, offset-free comb spectrum spanning more than one octave, from 0.14 to 0.325 THz at room temperature. Moreover, we show that the repetition frequency of the TFC can be tuned from 0.9 to 50 GHz by the injected RF signal, and a 120 Gbps (16 Gbps per channel) data rate can be achieved in the TFC-based high-speed single-channel (multi-channel) wireless communications.**

# Introduction

Frequency combs are high-coherence sources distinguished by a spectrum of regularly spaced, phase-locked spectral lines. Their exceptional attributes—high frequency stability, low phase noise, and precise frequency scaling—have established them as indispensable tools in fundamental time metrology, high-precision spectroscopy, high-resolution sensing, and high-speed wireless communication[1-4]. However, research efforts have predominantly focused on optical frequency combs, leaving terahertz (THz) frequency combs (TFCs) comparatively underdeveloped. Realization of a compact, tunable, and broadband TFC operating at room temperature under electrical pumping has thus far remained a longstanding and formidable challenge. This is primarily due to the absence of sufficiently advanced THz sources. Consequently, the adoption of this technology for widespread THz applications, such as molecular spectroscopy and wireless communications, has been severely hindered.

To date, several techniques have been developed for realizing compact TFCs. The first involves mode-locked THz quantum cascade lasers (QCLs)[5-6]. However, these typically require cryogenic cooling, limiting their practical use. Although recent advances in difference-frequency generation (DFG) have enabled room-temperature operation, it remains extremely challenging for these combs to generate frequencies below 2.2 THz, and they suffer from narrow bandwidths[7]. A second approach leverages Si complementary metal-oxide-semiconductor (CMOS) technology, but these combs often require external digital triggering[8-10], have limited bandwidth[11-15], and cannot efficiently generate combs above 1.1 THz. A third technique utilizes direct current (DC) driven resonant tunneling diodes (RTDs), which operate at room temperature and span a broad frequency range from 0.1 to 1.98 THz[16-18]. While the first RTD-based TFC has been demonstrated via passive mode-locking[19], this method relies on cumbersome optical setups and is sensitive to external THz wave feedback, hindering the development of compact and tunable TFC sources.

In this work, we present the first electrically pumped, broadband, compact, and tunable TFC based on an actively mode-locked RTD operating at room temperature. Under strong radio frequency (RF) injection, the RTD generates a TFC spanning 0.14 to 0.325 THz without external THz wave feedback. The comb's repetition frequency (mode spacing) can be tuned via the injection RF frequency, and the measured phase noise of the beat note is below -100 dBc/Hz at a 100 Hz offset. Furthermore, we experimentally demonstrate a coherent transmitter using this RTD-based TFC, achieving a record-breaking single-channel data rate of 120 Gbps with 32-QAM for RTD-based wireless communication. We also establish a three-channel link with a data rate of 16 Gbps per channel using 16-QAM, validating its multi-channel capability for high-speed data routing. These results represent a significant advancement in coherent THz sources, opening new avenues for future high-speed wireless communication.

**Fundamental principle and architecture of the TFC source.**

Figure 1a illustrates the design of the TFC source, which is based on an actively mode-locked RTD oscillator (black dashed rectangle) integrated within a WR-5 hollow metallic waveguide. The design incorporates a cylindrical aperture to accommodate a bias line, enabling simultaneous DC bias and RF signal injection. The RTD oscillator comprises a split-ring resonator, a shunt resistor-terminated slot resonator (Fig. 1b), and an active region (Fig. 1c). Together, the split-ring and slot resonators serve as a coupler that efficiently extracts the TFC signal from the RTD oscillator into the hollow metallic waveguide. The active region consists of a sandwiched AlAs/InGaAs/AlAs heterostructure (red and green regions) grown on a semi-insulating (SI) Indium Phosphide (InP) substrate. This configuration creates a double-barrier quantum well structure, essential for the resonant tunneling of electrons[20] under an applied bias voltage (Fig. 1d).

The resonant tunneling effect generates a negative differential conductance (NDC) in the current-voltage (*I-V*) characteristic of the RTD, providing strong nonlinearity and gain in the THz frequency range[21]. When a strong RF signal is injected into the RTD oscillator, this inherent nonlinearity and gain excite and amplify multiple modes at integer multiples of the RF frequency $f_{RF}$. Consequently, the RTD becomes actively mode-locked, forming a broadband, offset-free TFC with a repetition frequency $f_{rep}$ precisely equal to $f_{RF}$. As a result, $f_{rep}$ can be tuned by varying $f_{RF}$, and the frequency of each comb line $f_n$ is given by:

$$f_n = nf_{RF}$$

The number and intensity of the comb modes are inversely correlated, a relationship primarily controlled by the $f_{RF}$ under constant injection power $P_{RF}$, as depicted in the right panel of Fig. 1a. Specifically, a higher $f_{RF}$ yields fewer but more intense comb modes, whereas a lower $f_{RF}$ produces more numerous but less intense comb modes. This trade-off stems from the approximately constant total available energy being redistributed among the comb modes.

**Fabrication of the TFC source**

We now fabricate the TFC source. Figure 2a shows the photograph of the fabricated device, which integrates an RTD oscillator within a WR-5 hollow metallic waveguide (see details of the coupling between the RTD oscillator and the metallic waveguide in Methods and Extended Data Fig. 1). A 1.85-mm coaxial connector, serving as the bias line, is directly inserted into the waveguide and connected to the RTD oscillator via wire bonding (see details of the wire bonding in Methods and Extended Data Fig. 2). The 100-μm-thick RTD chip (green dashed rectangle) is positioned at the center of the waveguide, between its upper and lower halves (Fig. 2b). Optical microscope and scanning electron microscope (SEM) images of the RTD oscillator are presented in Fig. 2c, d, respectively, with their corresponding equivalent circuit diagrams provided as insets (see detailed geometrical parameters, epitaxial structure, and fabrication process in Methods and Extended Data Figs. 3, 4).

The oscillator circuit consists of two series circuits formed by the slot and split-ring resonators (inset of Fig. 2c), connected in parallel with the RTD's intrinsic conductance $-G_{RTD}$ and capacitance $C_{RTD}$ (inset of Fig. 2d). The inductances from the resonator's short ($L_s$) and long ($L_l$) segments, together with $C_{RTD}$ and resonators' parasitic capacitance $C_p$, form an oscillator that supports the fundamental mode of free-running THz oscillation. The radiation resistance $R_{rad}$, in parallel with RTD, determines the power radiated from the oscillator. The $-G_{RTD}$ was extracted from the measured *I–V* characteristic (red dots in Fig. 2e), agreeing well with the simulation results (blue line in Fig. 2e). The bias voltage required to enter the NDC region (grey region) shifts from approximately 0.6 to 1.9 V due to the influence of parallel and series resistors near the RTD (see Extended Data Fig. 5b-d for the detailed DC circuit and *I-V* curve change). Free-running THz oscillation occurs when $-G_{RTD}$ is sufficient to overcome the combined losses from the short-slot resistance ($R_s$), long-slot resistance ($R_l$), shunt resistance ($R_{shunt}$), and $R_{rad}$, simultaneously providing the gain necessary to drive the TFC.

To analyze the gain and loss conditions, we performed electromagnetic simulations to calculate the total admittance $Y$ of the split-ring and slot resonators (blue lines) and the RTD (red lines) as a function of frequency (Fig. 2f). Both the real (solid lines) and imaginary (dashed lines) parts of the admittance are presented. The free-running oscillation frequency is determined by the intersection point of $Im(Y)$ and $j\omega C_{RTD}$ (black dashed circle). A positive effective gain, given by $G_{RTD}-Re(Y) > 0$, confirms the presence of sufficient gain for TFC generation.

Figure 2g shows the measured spectrum of the fundamental free-running oscillation mode at 0.188 THz, with an output power of −12 dBm (see details of power measurement in Methods and Extended Data Fig. 6), confirming the single-mode operation. However, the spectrum exhibits a relatively broad linewidth with significant phase noise. As this fundamental mode acts as a parasitic mode during TFC generation, we employ an active mode-locking technique based on RF injection to effectively suppress it.

To explore the physical mechanism of active mode-locking in the RTD oscillator, we employ the equivalent circuit model described above (insets in Fig. 2c, d) for circuit-level simulations. Theoretically, the harmonic balance method yields a broadband TFC spanning from 0.14 to 0.325 THz under high-power RF injection, as illustrated in Fig. 2h (see detailed simulation setting in Methods and Extended data Fig. 5a). Note that the influence of the metallic waveguide is omitted in the simulations to ensure generality across different device implementations. These results demonstrate that actively mode-locked RTD oscillators can support a broadband TFC.

**Experimental characterization of the TFC source.**

Next, we experimentally characterize the TFC source. The measurement setup is shown in Fig. 3a. A signal generator (R&S SMW200A) and a DC power supply (Keysight B2901BL) were used to drive the packaged RTD oscillator to generate the TFC. The output signal passes through a frequency extender (Ceyear 82407RA) and is measured by a spectrum analyzer (R&S FSW85). As shown in Fig. 3b, under an injected RF signal of $f_{RF}$ = 0.918 GHz and $P_{RF}$ = 18 dBm, the measured power spectral density (PSD) reveals a broadband TFC spanning 0.14–0.19 THz (see Methods and Extended Data Fig. 7 for the additional spectra covering 0.19–0.325 THz and under different $f_{RF}$ injections). The measured mode spacing matches $f_{RF}$ exactly and can be continuously tuned from 0.9 GHz to approximately 50 GHz—the upper limit accessible in the experiment. Magnified views of the spectra around 0.151 THz (orange line), 0.167 THz (cyan line), and 0.185 THz (blue line) are presented in Fig. 3c. Compared to the several-MHz linewidth of the free-running RTD oscillator (Fig. 2g), the linewidth under active mode-locking is reduced to 1 Hz. It should be noted that this 1 Hz value does not represent the intrinsic linewidth limit of the oscillator but is limited by the resolution of the measurement system. The uniform mode spacing and narrow linewidth confirm the successful active mode-locking of the RTD oscillator via RF injection, resulting in the generation of a broadband TFC.

To further evaluate the coherence of the comb modes under active mode-locking, phase noise measurements were performed. Since the comb modes are equally spaced with a mode spacing of $f_{RF}$, their homodyne beat notes coincide at this frequency. Therefore, the phase noise measured at $f_{RF}$ reflects the combined contribution from all beat signals and serves as an indicator of the overall comb coherence. Accordingly, the comb signal was detected using a Schottky barrier diode, and its phase noise at $f_{RF}$ was measured with a phase noise analyzer (R&S FSWP50). Figure 3d compares the phase noise of the comb state (red curve, $P_{RF}$ = 18 dBm) and the modulation state (blue curve, $P_{RF}$ = −10 dBm) as a function of frequency offset. A phase noise reduction exceeding 80 dB is observed in the comb state (see Extended Data Fig. 8 for the additional phase noise measurement results under different $f_{RF}$ and offset frequencies). The frequency stability (blue line) and power stability (red line) of the TFC are shown in Fig. 3e. The comb frequency remained stable over one hour of continuous operation, with power variations within about 0.2 dB. These results demonstrate that the RTD-based TFC functions as a stable, low-phase-noise coherent THz source.

To achieve active mode-locking from the fundamental mode, a sufficiently high RF injection power is required. Under free-running operation (i.e., without RF injection), the RTD oscillator emits only a fundamental mode at 0.192 THz, as shown by the bottom cyan line in Fig. 3f. When a weak RF signal ($P_{RF}$ = −5.6 dBm, $f_{RF}$ = 20 GHz) is injected, two modulation sidebands appear around the fundamental mode (blue line). As $P_{RF}$ increases from −5.6 dBm (blue line) to 1.4 dBm (yellow line), the fundamental mode becomes progressively unstable, and the modulated sidebands are suppressed. In this transition regime, comb modes begin to emerge, though their power remains relatively low. With further increase in the RF power to 11.8 dBm (orange line), the fundamental mode is strongly suppressed, while the comb mode power rises significantly—albeit with residual noise still visible near the comb lines. Finally, at $P_{RF}$ = 13.4 dBm (red line), the residual noise is suppressed, and the fundamental mode becomes fully locked to the comb modes, yielding a clean and well-defined comb spectrum. It should be noted that the RF power required for comb generation is closely linked to the nonlinearity of the NDC region, which in turn depends on the bias voltage. We observed that the NDC nonlinearity is strongest in its initial region, around bias voltages of 1.9–2.1 V. In this range, the comb can be generated with relatively low RF power, while still achieving higher comb-mode power compared to other bias conditions (see the phase diagram of measured comb power as a function of RF power and bias voltage in Extended Data Fig. 9).

Compared to TFCs realized using other techniques—such as CMOS (green circles), QCL (blue circles), DFG-QCL (cyan circles), and passively mode-locked RTD (red circle)—the actively mode-locked RTD-based TFC (red star) offers superior performance in both comb emission bandwidth (defined as the ratio of bandwidth to center frequency) and operating temperature, as shown in Fig. 3g. While QCL-based combs (blue circles) typically require cryogenic cooling, and room-temperature DFG-QCL combs exhibit limited bandwidth, other room-temperature alternatives like CMOS and passively mode-locked RTD combs also face inherent bandwidth constraints (see detailed comparison in Extended Data Table 2). By implementing active mode-locking in the RTD platform, we achieve a markedly expanded emission bandwidth under room-temperature conditions. This broadband comb generation facilitates the more efficient use of spectral resources, making it particularly suitable for high-capacity THz wireless communication systems.

**TFC-based high-speed THz wireless communication**

Finally, we demonstrate high-speed THz wireless communication using the TFC at the transmitter side, implementing both a single-channel link with high data rates and a multi-channel link that exploits multiple comb modes for parallel data transmission. The schematic of the TFC-based wireless communication system is shown in Fig. 4a. The TFC signal is first amplified by a THz low-noise amplifier (LNA) and then mixed with the baseband signal using a fundamental mixer (FM), which simultaneously upconverts the baseband data onto all comb modes. The modulated signal is transmitted wirelessly via a pair of THz horn antennas. At the receiver side, the incoming signal is amplified by another THz LNA and down-converted using a subharmonic mixer (SHM). The local oscillator (LO) required for the SHM is generated by frequency multiplication of an RF source ($LO_m$), where the relationship $LO_m \times 12 = f_n$ is used for homodyne detection in this setup. After final amplification by an LNA, the baseband data can be demodulated and received by different users (see detailed parameters of the wireless communication setting in Methods). The experimental setup for TFC-based wireless communication over a 5 cm link is shown in Fig. 4b.

Since the baseband signal is simultaneously up-converted to all comb modes, its maximum usable bandwidth is limited to approximately half of the comb mode spacing ($f_{RF}$) to prevent inter-channel crosstalk. To achieve a high single-channel data rate, we selected $f_{RF}$ = 49.5 GHz and set the baseband bandwidth to 24 GHz, ensuring sufficient guard band to suppress inter-channel interference. Leveraging the low phase noise of the actively mode-locked RTD-based TFC, we successfully transmitted a 24 Gbaud 32-QAM signal over a wireless link. Using an offline digital signal processing (DSP) chain (see details in Methods) to demodulate the signal and compensate for nonlinear impairments, we recovered a clear constellation diagram (Fig. 4c) with a bit error rate (BER) of $3\times10^{-2}$—close to the forward error correction limit—and achieved a data rate of 120 Gbps, the highest reported for any RTD-based THz wireless communication system.

To demonstrate multi-channel THz wireless communication using distinct comb modes, we implemented a TFC-based system with $f_{RF}$ = 17.7 GHz and selected three carriers—$f_9$ = 0.159 THz, $f_{10}$ = 0.177 THz, and $f_{11}$ = 0.195 THz—as independent transmission channels. Figure 4d shows the recovered 4 Gbaud 16-QAM constellation diagrams for all three channels, each achieving a data rate of 16 Gbps, which is sufficient for real-time, uncompressed 4K video transmission. The small observed variation in BER among channels suggests the feasibility of synchronized data transmission across multiple comb modes, indicating the potential of RTD-based TFCs to serve as multi-channel routing devices in future THz wireless communications.

## Discussion

In conclusion, we have experimentally demonstrated, for the first time, an electrically pumped, broadband, room-temperature, tunable, and compact TFC spanning the frequency range from 0.14 to 0.325 THz, based on an actively mode-locked RTD oscillator integrated into a hollow metallic waveguide—a configuration that, to our knowledge, has never been reported before. Active mode locking is achieved by injecting a strong RF signal into the RTD oscillator, and the comb mode spacing (repetition frequency) can be continuously tuned from 0.9 GHz to approximately 50 GHz by varying the RF frequency. This stands in contrast to the actively mode-locked QCL-based combs[5,24,26,31,33-36], whose repetition rates are constrained by the dimensions of their external cavity. Given that RTD oscillations have been demonstrated up to 1.98 THz (with a potential extension to 2.77 THz[37]), this work paves the way for future ultra-broadband TFC generation based on RTD technology. Furthermore, we have demonstrated high-speed THz wireless communication using the TFC as a multi-carrier source, achieving a single-channel data rate of 120 Gbps and a multi-channel transmission rate of 16 Gbps per channel, supporting applications such as real-time, uncompressed 4K video transmission. These results underscore the potential of RTD-based TFCs as compact, broadband, and tunable THz sources, positioning them as promising candidates for future integrated THz systems and high-capacity wireless links.

Looking forward, further increasing the output power and operational bandwidth of RTD-based TFCs represents an important research direction. Given that the WR-5 hollow waveguide primarily operates in the frequency range of approximately 0.14–0.22 THz, it is notable that replacing hollow waveguides with broadband packaging structures (e.g., dielectric lenses) can enable a TFC signal with a significantly wider bandwidth. Although spectral and phase noise measurements confirm the ultra-low phase noise of the generated comb, constellation analysis reveals that the wireless transmission performance is currently limited by the output power of TFC. To address this, future efforts should focus on developing RTD oscillators with lower loss and higher gain, thereby enabling high-power, broadband TFC sources suitable for next-generation THz communication and other emerging applications.

## Methods

**Coupling between RTD and WR-5 waveguide.** To investigate the mechanism of TFC extracted from the RTD oscillator coupled into the WR-5 waveguide, we performed a three-dimensional (3D) electromagnetic simulation. In the simulation, we placed a wave port at the end of the waveguide and replaced the RTD with a lumped port, as shown in Extended Data Fig. 1a-b. Extended Data Fig. 1c showed the calculated transmission coefficient $S_{21}$ and the coupling efficiency. The coupling efficiency is defined as

$$\eta = \frac{|S_{21}|^2}{1 - |S_{11}|^2} \tag{1}$$

, which represents the power extracted from the RTD oscillator and coupled into the waveguide. The simulation indicates a coupling efficiency of approximately 40% near 0.18 THz, resulting from relative effective impedance matching between the RTD oscillator and the waveguide. Conversely, a significant impedance mismatch occurs around 0.19 THz, leading to poor coupling. This is the primary reason for the degradation of the measured TFC power at frequencies above 0.19 THz. Extended Data Fig. 1d-1i illustrate the simulated electric field (E-field) distributions in the x-z, x-y, and y-z planes for the RTD oscillator inside the waveguide (Extended Data Fig. 1d-f) and the RTD oscillator itself (Extended Data Fig. 1g-i), respectively. The intense E-field generated by the RTD oscillator couples into the waveguide primarily through the long part of the slot resonator. The E-field leaked from the short part of the slot resonator propagates to the left-hand end of the waveguide, located 0.8 mm from the RTD chip. This waveguide end functions as a metal reflector, causing the reflected wave to travel back in the right-hand direction. The reflected wave also travels in the right-hand direction. The presence of a traveling wave propagating along the waveguide confirms the effective extraction of the TFC power from the RTD oscillator to the waveguide. A magnified view of the E-field in the vicinity of the RTD oscillator reveals a strong field concentration, indicating the resonant behavior of the slot and split-ring resonators.

**Packaging and fabrication process.** The WR-5 waveguide package was fabricated by Computer Numerical Control (CNC) milling in two parts. As shown in Fig. 2b, the RTD chip is mounted inside the upper half of the package. The magnified microscope picture showing the details of the RTD chip packaged to the waveguide is shown in Extended Data Fig. 2. A 1.85-mm RF coaxial connector is inserted through a hole positioned near the chip location, appearing as the inner conductor of the connector in Extended Data Fig. 2. After fixing the connector with two screws, it shares a common ground with the waveguide package.

The RTD chip was polished from 600 to 100 μm and diced into 300×900 μm dies. To mount the RTD, a small amount of wax was first dipped into the mounting hole. After cooling, the RTD chip was placed on the wax using tweezers. The wax was then reheated and cooled again to firmly attach the chip to the package. Wire bonding was carried out to connect the inner conductor of the connector to the RTD. A second bonding wire was taken directly from the waveguide package to the RTD. Since the package and connector share the same ground, this configuration allows both DC biasing and RF signal injection through the connector.

Extended Data Fig. 3 depicts the detailed geometrical parameters of the RTD oscillator and the epitaxial layer structure of the RTD wafer. The short slot, long slot, and perimeter of the split ring are 60, 100, and 60 μm, respectively. The split ring enhances the concentration of the electric field within the slot resonator, thereby reducing losses[38]. The incorporation of Al element into the InGaAs layer, grown on the n$^+$InP layer, results in a reduction of the conduction band energy level on the emitter side of the RTD[39]. This design allows the device to enter the negative differential conductance (NDC) region at a low bias voltage. The 10 nm n$^+$-InP layer serves as an etch stop layer[40] during the wet etching process for the subsequent fabrication steps. Extended Data Fig. 4 describes the fabrication process of the RTD device. A double-layer resist consisting of LOR-2A and PMGI was coated, and photolithography was employed to define the electrode pattern. After depositing Ti/Pd/Au with a thickness of 20/20/200 nm, we performed the lift-off process to form the electrode. The Ti/Pd/Au provides good ohmic contact with n$^+$-InGaAs. For the subsequent etching step, we used AZ9260 as the dry etching mask. We applied a gas mixture of $BCl_3$:$Cl_2$:Ar = 15:5:8 for the ICP-RIE process to remove the n$^+$-InGaAs inside the slot and outside the DC pads. The shunt resistors were thereby formed, and different RTD elements were isolated. RZJ304-10 was then employed as a protection mask for the wet-etching process used to form the air bridge. Due to the isotropic etching with etchant ($H_3PO_4$:$H_2O_2$:$H_2O$ = 1:1:80), the n$^+$-InGaAs beneath the air bridge was removed laterally, resulting in the formation of the suspended air bridge. After removing the RZJ304-10 protection mask, a final wet-etching step was carried out to remove the semiconductor material near the RTD electrodes, thereby forming the target RTD mesa.

**Circuit simulation.** We modeled the RTD oscillator under TFC conditions using the equivalent circuit shown in Extended Data Fig. 5a. TFC generation requires strong RF injection, which we represent as an RF source with a 50 Ω source impedance. The source is in parallel with RTD, and its power is represented as $P_{RF}$. The oscillator circuit includes the RTD in parallel with its resonator load, a simplified model derived from the structure in Fig. 2c, radiation resistance $R_{rad}$, and resonators' parasitic capacitance $C_p$. The equivalent resonator resistance $R_{load}$, resonator inductance $L_{load}$, $R_{rad}$, and $C_p$ are approximately 12.7 Ω, 31.4 pH, 460 Ω, and 5.8 fF, respectively. The RTD area $S_{RTD}$ is 2.7 μm$^2$. We modeled the RTD with a linear capacitance $C_{RTD} = S_{RTD} \times 11$ fF/μm$^2$, and a nonlinear voltage-dependent conductance $G_{RTD}(V) = I_{RTD}(V)/V$. To extract the RTD current $I_{RTD}(V)$, we performed DC measurements of RTD. In the DC measurement setup (Extended Data Fig. 5b), parasitic components, including a shunt resistor ($R_{shunt} \approx 9$ Ω) and series resistance ($R_{series} \approx 15$ Ω) from the bias tee, obscure the intrinsic RTD characteristics. Consequently, the measured NDC region shifted to ~1.9–2.6 V (Extended Data Fig. 5d, blue line). To extract the I-V characteristics required for the high-frequency simulation in Extended Data Fig. 5a, we fabricated and measured a separate RTD device lacking the $R_{shunt}$ and $R_{series}$. The test setup in Extended Data Fig. 5c minimizes parasitic resistances and provides a relatively accurate intrinsic I-V curve, as shown in Extended Data Fig. 5d (red line), revealing the NDC region between ~0.6 V and 1.1 V. The nonlinear current $I_{RTD}(V)$ used in the model was subsequently derived as[19,21]:

$$I_{RTD}(V) = S_{RTD}\left(C_1 V\{\tan^{-1} C_2(V - V_1) - \tan^{-1} C_2(V - V_2)\} + C_3 V^i\right) \quad (2)$$

The parameters of Equation 2 are shown in the Extended Data Table 1. We simulated the TFC circuit using the harmonic balance method with $f_{RF}$ = 0.918 GHz, $P_{RF}$ = 10 dBm, and 0.7 V bias. Taking the voltage across $R_{rad}$ as the output, we observed the TFC generation and the corresponding spectrum as presented in Fig. 2h.

**Measurement setup of output power, comb, and phase noise.** To characterize the output power of the RTD oscillator, we connected the module to an Erickson PM5 power meter through a WR-5 waveguide, as shown in Extended Data Fig. 6. In its free-running state at 0.188 THz, the module was −12 dBm. A THz LNA then amplified this signal to 1.3 dBm. To induce TFC operation, we injected a 20 GHz, 13 dBm RF signal. The module's output power was −19.1 dBm, and it was amplified by the THz LNA to −2.5 dBm.

We connected the packaged RTD oscillator with a bias tee for DC supply and RF injection to achieve the TFC generation. We down-converted the TFC using a frequency extender and fed it into the spectrum analyzer for TFC measurement. We directly connect the RTD oscillator and frequency extender to the waveguide. We use two frequency extenders (WR-5 & WR-3 waveguide types) covering the frequency ranges of 0.14–0.22 and 0.22–0.325 THz to measure the TFC. Notably, the generated comb modes persist at frequencies exceeding 0.3 THz, significantly surpassing the fundamental frequency of the RTD oscillator. This finding shows the feasibility of generating TFCs far beyond the RTD oscillator's fundamental mode. We also measured spectra under different $f_{RF}$ in Extended Data Fig. 7, showing that the frequency comb is tunable. During the measurements, we observed weak spectral lines between the comb modes, which appeared as measurement noise. These spurious components were effectively suppressed when the resolution bandwidth (RBW) and span of the spectrum analyzer were reduced to 100 kHz and 1 GHz.

The phase noise measurement diagram is shown in Extended Data Fig. 8a. We use a signal generator (R&S SMW200A) and a DC power supply (Keysight B2901BL) to drive the packaged RTD oscillator for the TFC generation. The Schottky barrier diode (SBD) functions as a direct detector to measure the repetition frequency $f_{RF}$, then an LNA is used to amplify the signal and pass it into a phase noise analyzer. The TFC module and SBD are connected directly through the WR-5 waveguide. The measured phase noise at 1 Hz (green circles), 100 Hz (blue circles), and 1 kHz (red circles) with different injection frequencies $f_{RF}$ is shown in Extended Data Fig. 8b.

**Sweeping bias voltage and $f_{RF}$.** To determine the conditions required for active mode-locking under RF injection, the RTD bias voltage and the injected RF power were systematically varied. The bias voltage was swept from 1.3 to 2.7 V, a range that encompasses the device's NDC region, which spans ~1.9–2.7 V. To prevent device breakdown, measurements were not conducted beyond this upper voltage limit. The device's operation can be roughly classified into free-running, modulation, unstable free-running, unstable comb, and comb states. These states are labeled in Extended Data Fig. 9b when $f_{RF}$ is 20 GHz. We neglect the free-running state because it does not exist when an RF signal is injected. We also neglect the unstable comb state because it is a transition state that only exists within the change of ~1–2 dBm. To visualize these operational regimes, the spectral power at 0.14, 0.16, 0.18, 0.2, and 0.22 THz was summed, with the total power plotted in Extended Data Fig. 9a. A low summed power is characteristic of the unstable free-running state, where modulation and comb states coexist (blue and cyan areas). A high summed power indicates stable comb generation (red and orange areas). Notably, comb states were observed at bias voltages far below the beginning of the NDC region. This is attributed to the large-signal RF injection, which induces a voltage swing sufficient to drive the RTD instantaneously into the NDC region. Furthermore, the strong power observed at the beginning of the NDC region is likely due to the pronounced nonlinearity of the device at that bias point.

**Wireless communication setup.** We use an arbitrary waveform generator to generate a pseudorandom binary sequence 15 baseband signal with a peak-to-peak voltage of 1 V. Before up-converting, we added a 3 dB attenuator to protect the FM. The two THz LNAs, each providing 16 dB gain within the 0.12–0.2 THz frequency range, are employed, setting the upper limit of the carrier frequency to 0.2 THz. The ×6 frequency multiplier, with an input frequency range of 12.5–18.33 GHz and an output frequency of 0.075–0.11 THz, is employed at the receiver side. Combining with SHM, the homodyne detection lower limit is 0.15 THz. For down-converted signals, a 38 dB LNA is used, operating at 10 MHz–50 GHz. Finally, a real-time oscilloscope was employed to capture and process the received data. To characterize the signal from the transmitter side, we replaced a frequency extender with the receiver components. The comb mode spectra with $f_{RF}$ = 49.5 and 17.7 GHz, used in the wireless communication demonstration, are shown in Extended Data Fig. 10a, b, respectively. To inject an RF signal at 49.5 GHz, we replace the signal generator used during measurement (with an upper limit of 44 GHz) with a Keysight E8257D PSG Analog Signal Generator. A signal-to-noise ratio (SNR) exceeding 20 dB was achieved for the up-converted baseband signal using $f_4$ = 0.198 THz as carrier frequency. This performance meets the minimum threshold required for a 120 Gbps wireless communication link employing a 32-QAM modulation scheme. The spectrum in Extended Data Fig. 10b confirms the feasibility of multi-channel communication by utilizing distinct comb modes as separate data carriers. The observed variance in the constellation diagrams in Fig. 4d is attributed to the different SNR using $f_9$ = 0.159, $f_{10}$ = 0.177, and $f_{11}$ = 0.195 THz in Extended Data Fig. 10b, respectively.

**Digital signal processing.** The single-carrier (SC) modulation is set up in our experimental systems, which is considered a spectral- and energy-efficient technology for THz wireless communications[41]. The main offline DSP flows are shown in Extended Data Fig. 11. At the Tx DSP module, SC-16QAM/32QAM signals are generated with randomly chosen symbol sequences of length $2^{16}$ and up-sampled at 2 samples per symbol. Subsequently, a raised-cosine (RC) filter with a roll-off factor of 0.1 is implemented for the pulse shaping. These shaping signals, after undergoing digital up-conversion (DUC) and resampling, are used to drive the mixer. At the Rx DSP module, received signals are initially processed by digital down-conversion (DDC) to obtain baseband signals and then passed through a low-pass filter to eliminate out-of-band noise. Following this, a series of advanced DSP modules, including timing phase recovery, linear/nonlinear channel equalization, and noise cancellation, are implemented to enhance the system performance. The details of the above algorithms are as follows:

The timing phase recovery based on Godard timing error detection (TED)[42] is applied to mitigate sampling frequency/phase errors, which is a popular frequency domain TED algorithm widely used in SC communication systems. The principle of Godard TED can be expressed as

$$\tau(k) = \frac{angle\left(\sum_{n=(k-1)N+1}^{(k-1)N+\frac{N}{2}} S(n)S^*\left(n+\frac{N}{2}\right)\right)}{2\pi f_s} \quad (3)$$

where $\tau$ represents the timing error, $k$ represents the block index, $N$ represents the length of the fast Fourier transform (FFT) block, $S$ represents the FFT of the input signals, $f_s$ represents the sample rate, and $angle$ represents the operation of measuring an angle.

The linear channel equalization based on a T/2-spaced radius-direction equalizer (RDE)[43] is implemented to mitigate inter-symbol interference (ISI), which typically results from multipath propagation, non-flat channel frequency responses, and standing wave effects. RDE is a kind of adaptive filtering algorithm with the mean square error of power as the cost function. The advantage of RDE is insensitive to carrier frequency and phase errors, which is conducive to modular programming. The implementation processes of the algorithm are as follows

$$s_{out}(n) = \boldsymbol{w}^H \boldsymbol{s_{in}}(n) \quad (4)$$

$$\boldsymbol{w}(n+1) = \boldsymbol{w}(n) + \mu e(n) s_{out}^*(n) \boldsymbol{s_{in}}(n) \quad (5)$$

$$e(n) = |\hat{s}_{out}(n)|^2 - |s_{out}(n)|^2 \quad (6)$$

where $s_{out}$ and $\hat{s}_{out}$ represent the output signal and the related ideal symbol, $\boldsymbol{s_{in}}$ represents the input signal vector of $1 \times N$, $\boldsymbol{w}$ represents the tap vector of $1 \times N$, $\mu$ represents the step of update, $e$ represents the error function, the superscript '*' and 'H' represent conjugate and conjugate transpose operation. Notably, N is set to 151 in this paper, which is sufficiently long to eliminate most of the ISIs.

The nonlinear channel equalization based on T-spaced $2 \times 2$ Volterra equalizer (VE)[44] is used to compensate nonlinear distortion, phase noises and residual ISIs. The classic adaptive Volterra equalization algorithm is combined with multiple input multiple output (MIMO) architecture to deal with the nonlinear effects of inter- and intra in-phase/quadrature (IQ) channel. The core equation of $2 \times 2$ VE is as follows

$$s_{out,I}(n) = \sum_{k=1}^{K} \sum_{q_1=0}^{N} \cdots \sum_{q_k=0}^{N} w_{II}(q_1,\ldots,q_k) \prod_{m=1}^{k} s_{in,I}(n-q_m)$$

$$+ \sum_{k=1}^{K} \sum_{q_1=0}^{N} \cdots \sum_{q_k=0}^{N} w_{QI}(q_1,\ldots,q_k) \prod_{m=1}^{k} s_{in,Q}(n-q_m) \quad (7)$$

$$s_{out,Q}(n) = \sum_{k=1}^{K} \sum_{q_1=0}^{N} \cdots \sum_{q_k=0}^{N} w_{IQ}(q_1,\ldots,q_k) \prod_{m=1}^{k} s_{in,I}(n-q_m)$$

$$+ \sum_{k=1}^{K} \sum_{q_1=0}^{N} \cdots \sum_{q_k=0}^{N} w_{QQ}(q_1,\ldots,q_k) \prod_{m=1}^{k} s_{in,Q}(n-q_m) \quad (8)$$

where $K$ represents the nonlinear order, $N$ represents the memory length, $w_{II/QI/IQ/QQ}$ represents the tap of MIMO filter, the subscript 'I' and 'Q' represent I and Q parts of the received complex signals. In this paper, a $2 \times 2$ third-order Volterra equalizer with a memory length of 6 is applied. Such a short filtering scheme aims to strike a balance between complexity and performance.

Noise cancellation (NC) based on error autocorrelation calculation[45] is implemented to mitigate the amplified colored noise caused by frequency-selective fading. Both pre- and post-cursor noise correlations are considered in the proposed scheme, and the related principle is as follows.

$$s_{out}(n) = s_{in}(n) - \sum_{k=-\frac{N}{2},\ k\neq 0}^{\frac{N}{2}} c_k e(n-k) \quad (8)$$

$$c_k = \frac{\sum_{m=1}^{M-k} e(m)e^*(m+k)}{M-k}, \qquad c_{-k} = c_k^* \quad (9)$$

$$e(n) = \hat{s}_{in}(n) - s_{in}(n) \quad (10)$$

where $N$ is the half-length of NC filter, $c_k$ represents the error autocorrelation parameter, $M$ represents the symbol length used to calculate autocorrelation. $N$ is set to 15 in this paper.

## Data availability

The data that support the findings of this study are available from the corresponding authors upon reasonable request.

## Acknowledgments

X.B.Y. acknowledges funding from the Mobile Information Networks National Science and Technology Major Project (grant no. 2025ZD1303200), Z.G. acknowledges funding from the National Key R&D Program of China (grant no. 2025YFA1412300).

## Author contributions

X.B.Y., Z.G., and X.F.T. initiated the project. F.F.H. performed the theoretical calculation. F.F.H., X.B.Y., and Z.G. designed the experiments. F.F.H. and X.B.Y. fabricated the samples and carried out the measurements. F.F.H., X.B.Y., Q.Z., Z.B.H., L.H.Z., W.C.L., J.P.D., and Z.G. analyzed the results. F.F.H., X.B.Y., and Z.G. wrote the manuscript with input from all authors. F.F.H., X.B.Y., Z.G., and X.F.T. revised the manuscript. X.B.Y., Z.G., and X.F.T. supervised the project.

## Competing interests

The authors declare no competing interests.

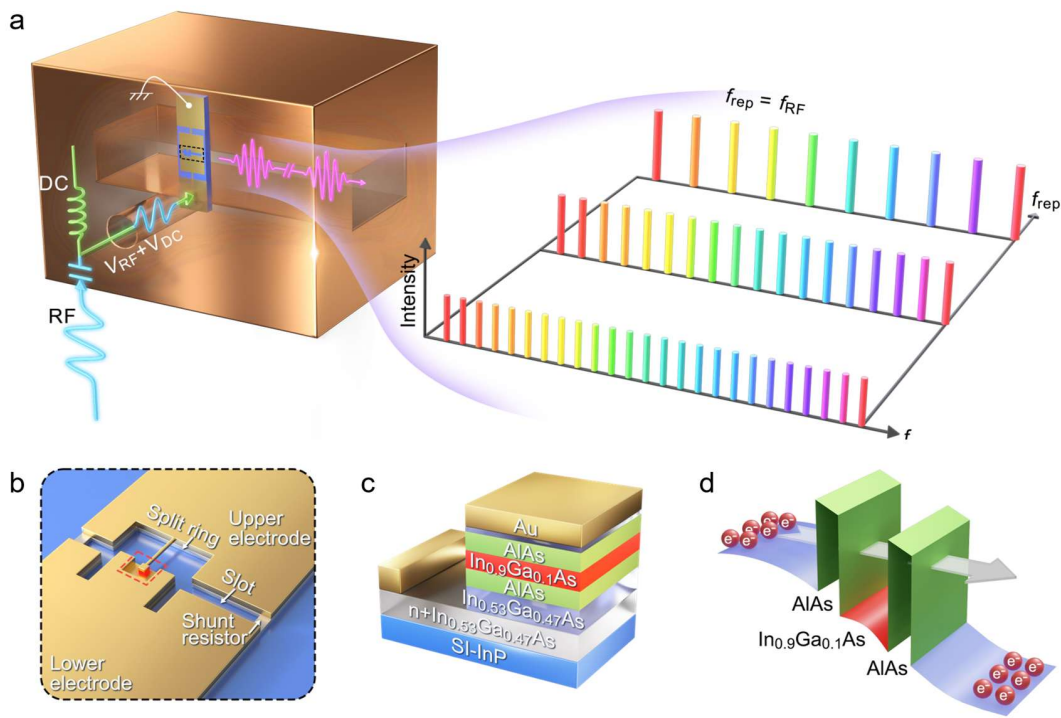

**Fig. 1 | TFC based on an actively mode-locked RTD integrated into a WR-5 hollow metallic waveguide. a,** Schematic of the TFC (left panel) with tunable mode spacing (right panel), which is generated by an actively mode-locked RTD under RF injection and DC bias. A cylindrical hole is designed to accommodate a 1.85-mm coaxial connector for supplying RF injection and DC bias. **b,** The RTD oscillator is composed of the active region (red square), split ring, and slot resonators terminated by shunt resistors. **c,** The active region (red and green colors) consists of a sandwiched AlAs/InGaAs/AlAs heterostructure on a SI-InP substrate. **d,** The band diagram of the active region features a double barrier and one quantum well structure, enabling the resonant tunneling of electrons.

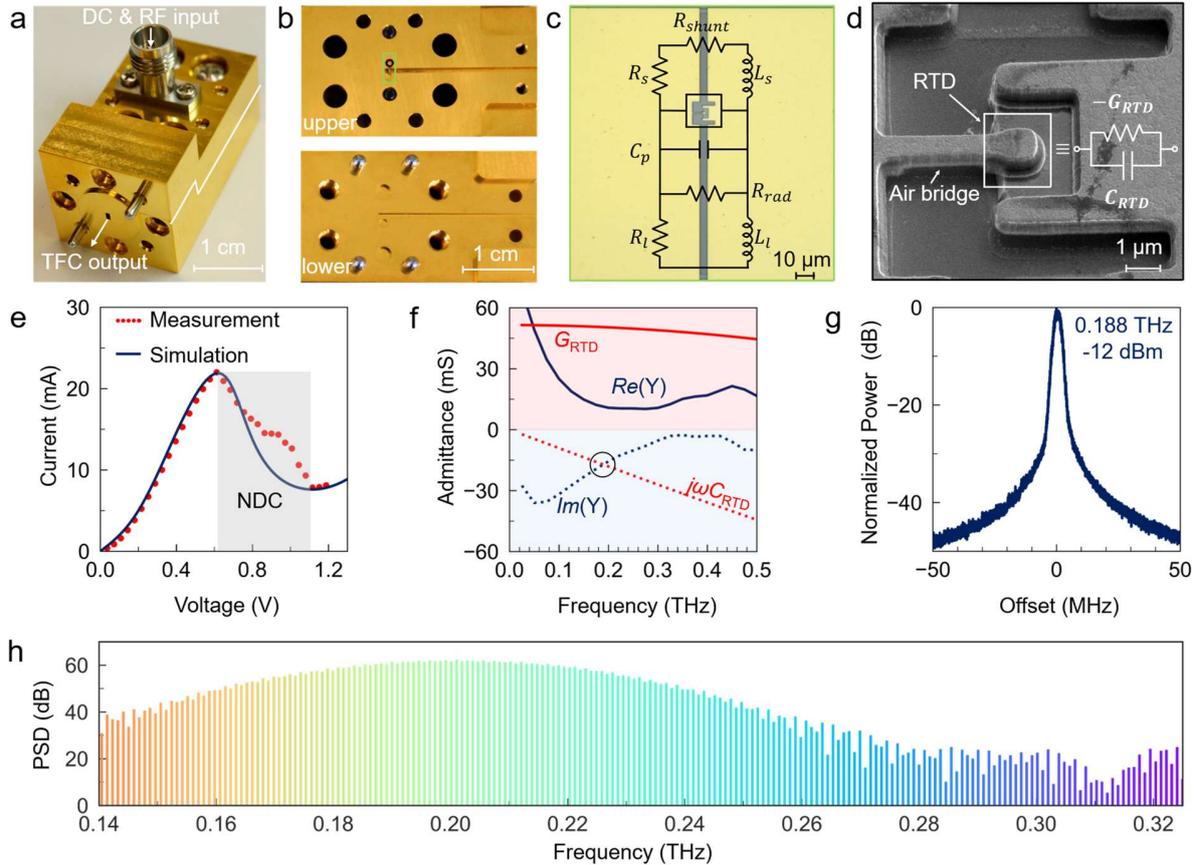

**Fig. 2 | Fabrication of the RTD-based TFC. a,** Photograph of the fabricated TFC. **b,** The upper and lower halves of the TFC. The green dashed rectangle represents the RTD integrated in the WR-5 hollow metallic waveguide. **c,** The optical microscope of the fabricated RTD oscillator chip (blue dashed square). The inset shows the equivalent circuit of the split-ring and slot resonators. **d,** The SEM photograph of the fabricated RTD oscillator chip and the equivalent circuit of the RTD. **e,** Measured (red dots) and simulated (blue line) $I$-$V$ characteristic of RTD under DC bias. The grey region represents the NDC region. **f,** Simulated total admittance $Y$ of the split-ring and slot resonators (blue solid and dotted lines) and simulated admittance of the RTD (red solid and dotted lines) at THz frequencies. The intersection between the imaginary part of $Y$ and $j\omega C_{rtd}$ (black dashed circle) indicates the free-running oscillation frequency of the RTD oscillator. **g,** Measured spectrum of the fundamental mode of the free-running RTD oscillator. **h,** Simulated TFC spectrum using the equivalent circuit of the RTD oscillator with 0.918 GHz RF injection.

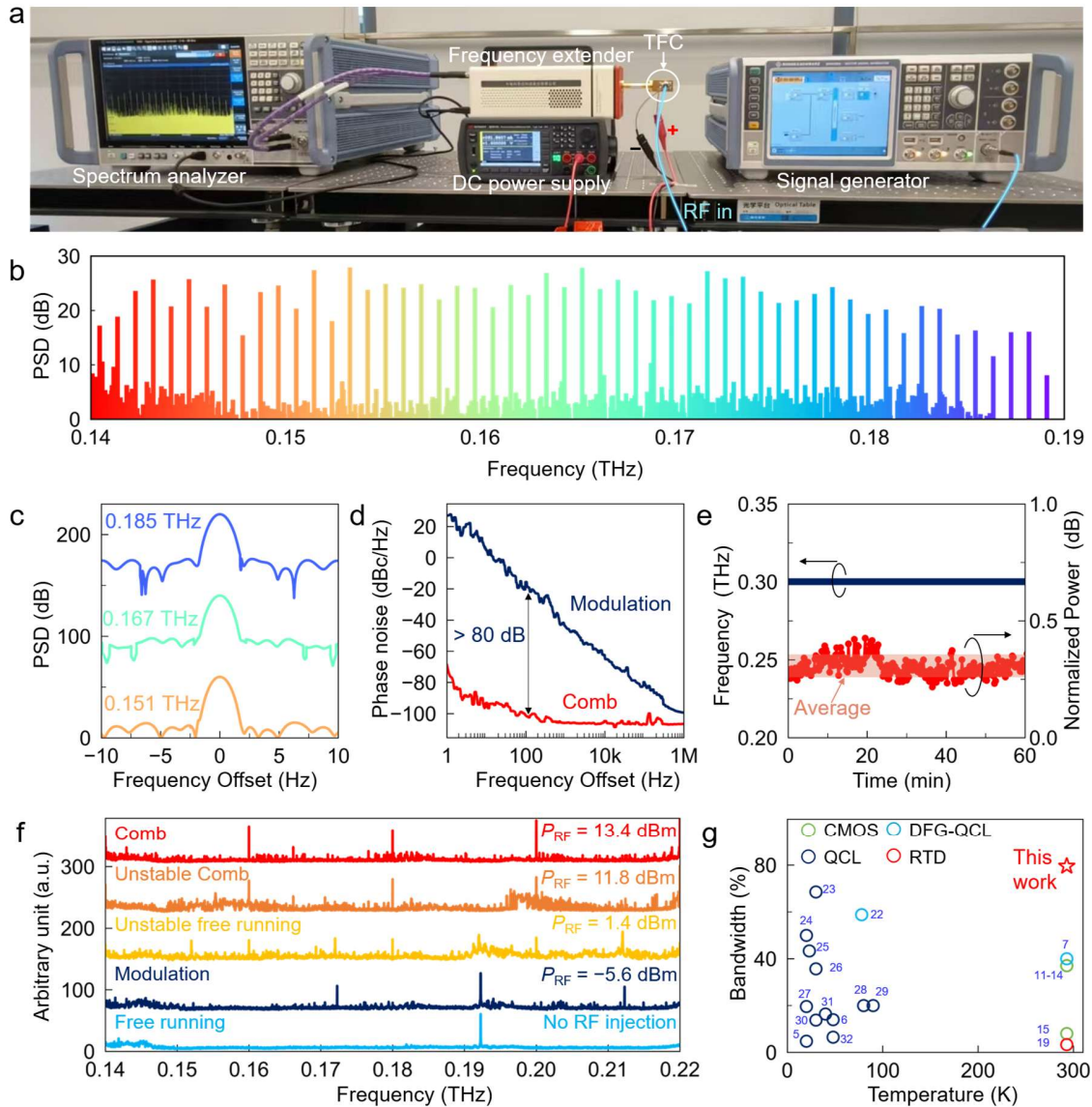

**Fig. 3 | Experimental characterization of the RTD-based TFC. a,** The experimental setup for characterizing the TFC. **b,** The measured PSD spectrum of the TFC spans from 0.14 to 0.19 THz with an injection frequency $f_{RF}$ = 0.918 GHz. **c,** The measured PSD spectra of individual TFC modes at 0.151, 0.167, and 0.185 THz, respectively, all have a 1 Hz linewidth. **d,** The measured phase noise of the TFC under comb (red line) and modulation (blue line) states. The phase noise decreases more than 80 dB when the TFC transitions from the modulation state with a lower injection power, $P_{RF}$ = −10 dBm, to the comb states with a higher injection power, $P_{RF}$ = 18 dBm. **e,** The measured frequency (blue line) and power (red line) fluctuation of the TFC modes over one hour. **f,** The evolution of RTD active mode-locking with increasing injected RF power. With gradually increasing RF injection power, the free-running RTD (cyan line) transitions from simple modulation (blue line) to unstable (yellow and orange lines) and comb (red line) state. **g,** Comparison of TFCs based on different techniques with different bandwidths and operating temperatures[5-7,11-15,19,22-32].

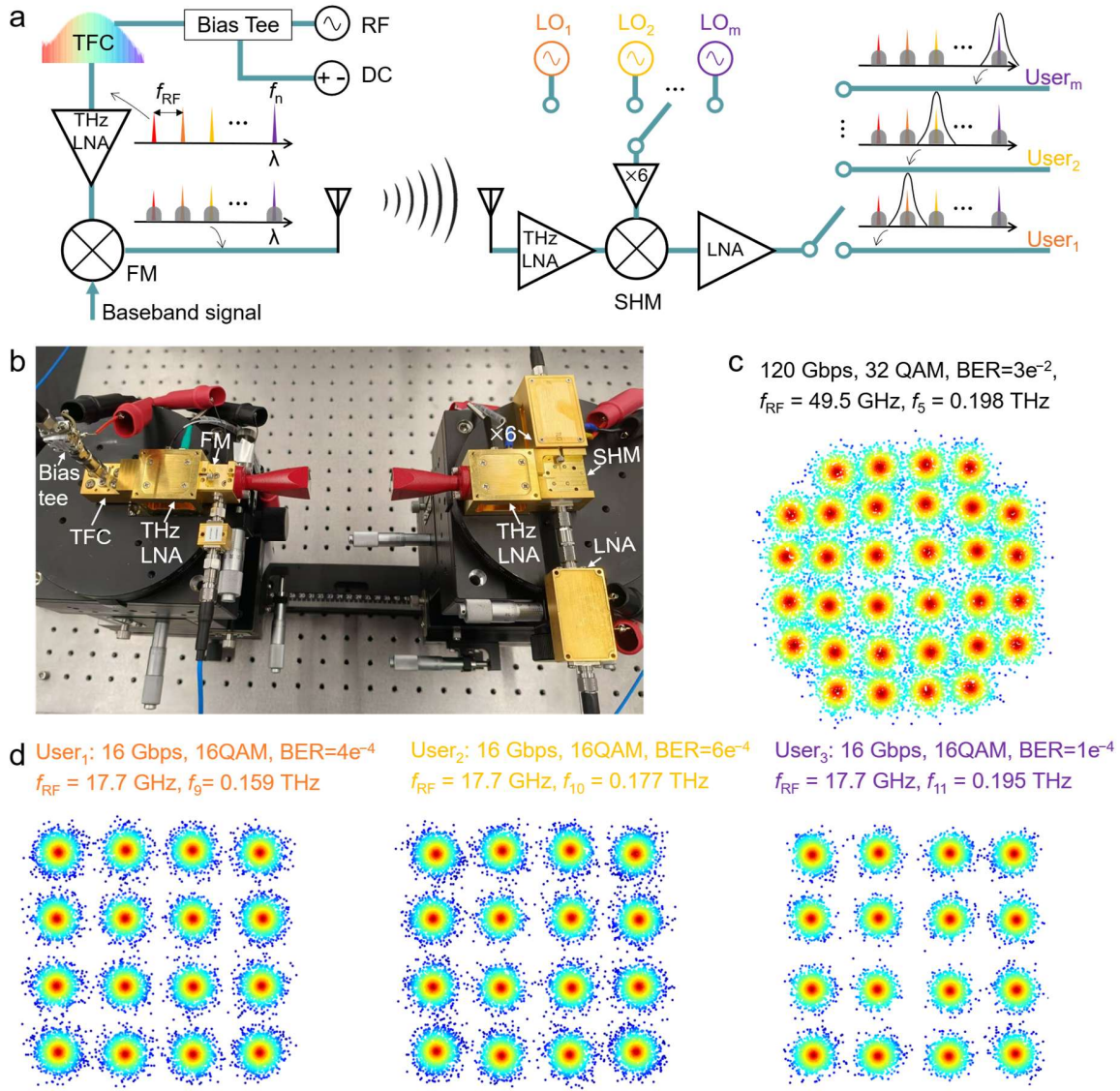

**Fig. 4 | TFC-based high-speed THz wireless communication. a,** Schematic of the TFC-based THz wireless communication. Identical baseband signals are up-converted onto different comb modes. By employing different LO signals, each baseband signal can be independently down-converted with homodyne detection. **b,** Experimental setup of the TFC-based THz wireless communication system. **c,** Measured constellation diagram of a single-channel THz wireless communication with a 120 Gbps data rate and a BER = $3e^{-2}$, using 32-QAM modulation, comb mode $f_5 = 0.198$ THz, and $f_{RF} = 49.5$ GHz, respectively. **d,** Measured constellation diagram of a multi-channel THz wireless communication with a 16 Gbps data rate using three different comb modes $f_9 = 0.159$ THz, $f_{10} = 0.177$ THz, and $f_{11} = 0.195$ THz, respectively.

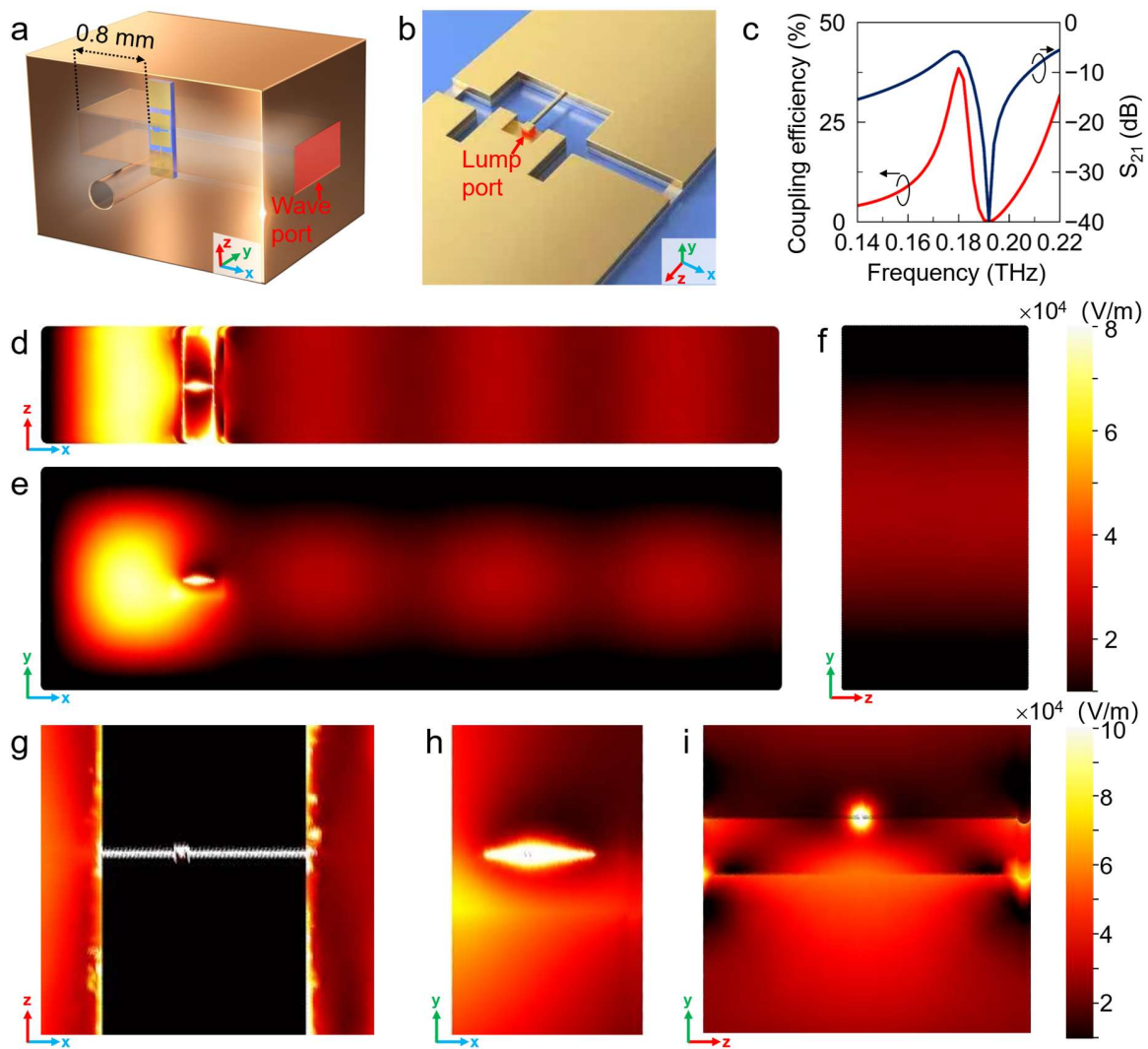

**Extended Data Fig. 1 | Electromagnetic simulations of the RTD in metallic waveguide. a,** Configuration of the wave port at the waveguide output. **b,** Lumped port assignment representing the RTD element. **c,** Calculated coupling efficiency and transmission coefficient S$_{21}$. **d-e,** Simulated electric field distributions in the x-z and x-y planes, respectively, illustrate the extraction of the electromagnetic wave from the RTD chip into the waveguide. **f,** Simulated electric field distribution in the y-z plane showing one of the wave crests inside the waveguide. **g-i,** Magnified electric field distributions in the x-z, x-y, and y-z planes of the RTD oscillator itself, respectively, indicate strong resonance and field confinement within the slot and split-ring resonators.

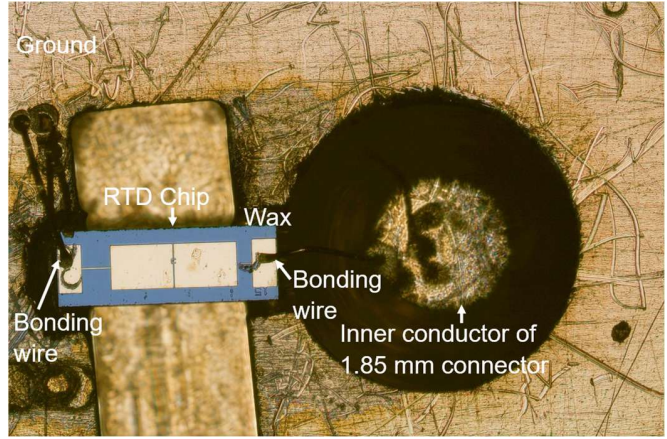

**Extended Data Fig. 2 | Detailed RTD packaging method.** The RTD chip is fixed within the WR-5 waveguide using wax. The inner conductor of a 1.85 mm coaxial connector is inserted through a cylindrical hole and positioned in close proximity to the RTD chip. We connected two bonding wires to the RTD chip. The right wire is connected to the inner conductor of the coaxial connector to establish the signal path. The left wire is directly connected to the package, serving as the signal ground.

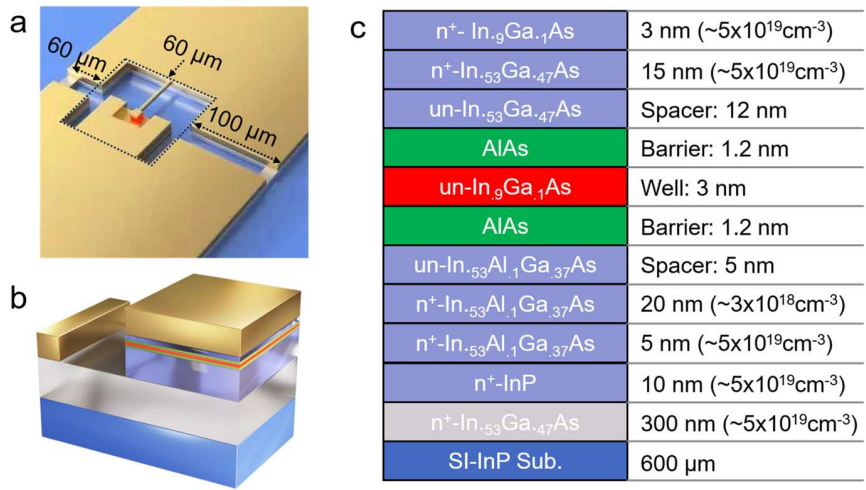

**Extended Data Fig. 3 | Detailed RTD resonator structure. a,** RTD oscillator composed of active region and resonator. The short slot, long slot, and perimeter of the split ring are 60, 100, and 60 μm, respectively. **b,** RTD active region in the vertical direction. **c,** Detailed epitaxy layer of RTD.

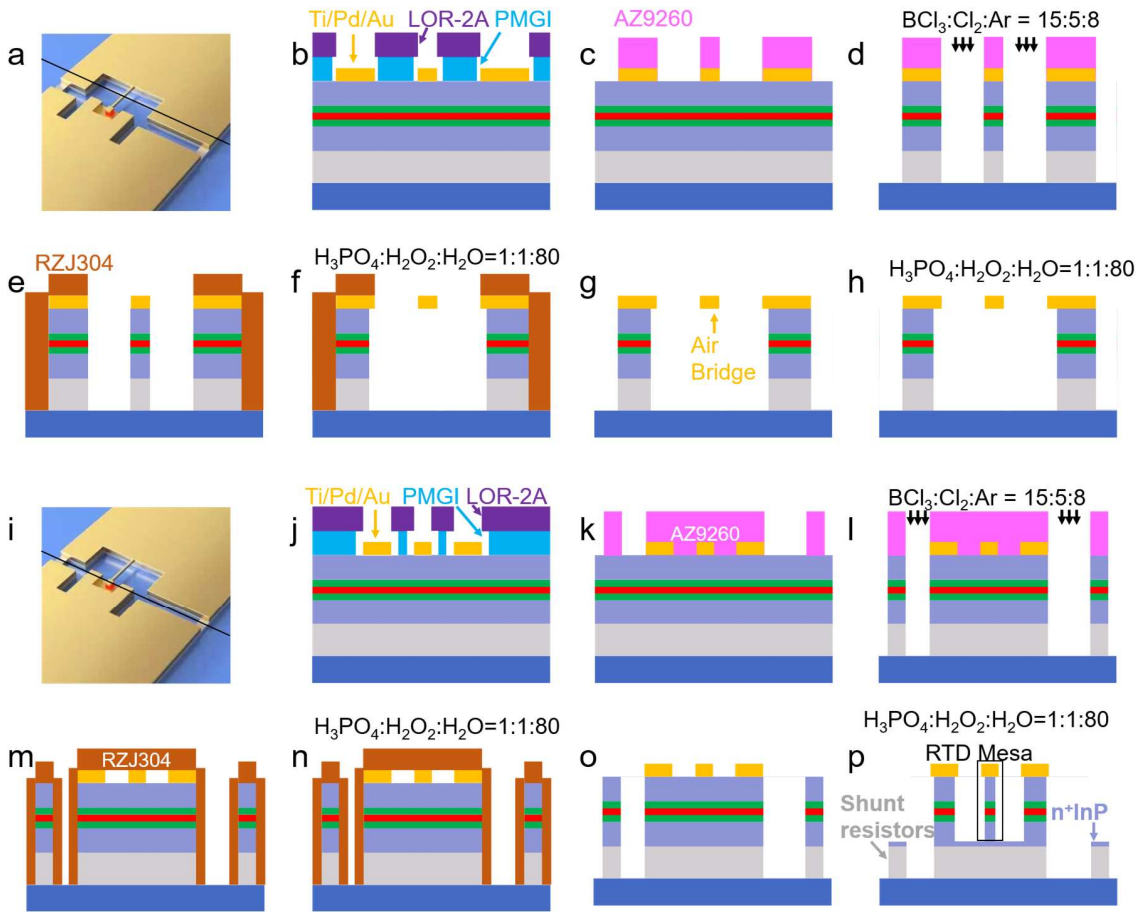

**Extended Data Fig. 4 | Detailed fabrication process. a,** Cross section of air bridge. **b,** LOR-2A and PMGI double layer lithography, and deposition of Ti/Pd/Au = 20/20/200 nm. **c,** AZ9260 as a resist mask for ICP-RIE. **d,** Gas mixture of $BCl_3:Cl_2:Ar = 15:5:8$ for semiconductor dry etching. **e,** RZJ304 is a resist mask for wet etching. **f,** An air bridge is formed using a wet etching process with an etchant ($H_3PO_4:H_2O_2:H_2O = 1:1:80$). Due to the isotropic direction of this chemical etch, the material layers beneath the metal are removed. This undercutting action results in the formation of the desired air bridge structure. **g,** Removal of RZJ304. **h,** Extra wet etching. **i,** Cross-section of RTD mesa and shunt resistors. **j,** LOR-2A and PMGI double-layer lithography, and deposition of Ti/Pd/Au = 20/20/200 nm. **k,** AZ9260 is a resist mask for ICP-RIE. **l,** Gas mixture of $BCl_3:Cl_2:Ar = 15:5:8$ for semiconductor dry etching. The shape of the shunt resistors was formed. **m,** RZJ304 as a resist mask for wet etching. **n,** The RTD part and the shunt resistors were protected from over-etching. **o,** Removal of RZJ304. **p,** An additional wet etching step is performed to fabricate the RTD mesa and the shunt resistors. The duration of this etch is precisely controlled to define the RTD mesa and achieve the target area. The etching process selectively stops at the $n^+$InP layer, which serves as an etch-stop and enables the formation of the shunt resistors.

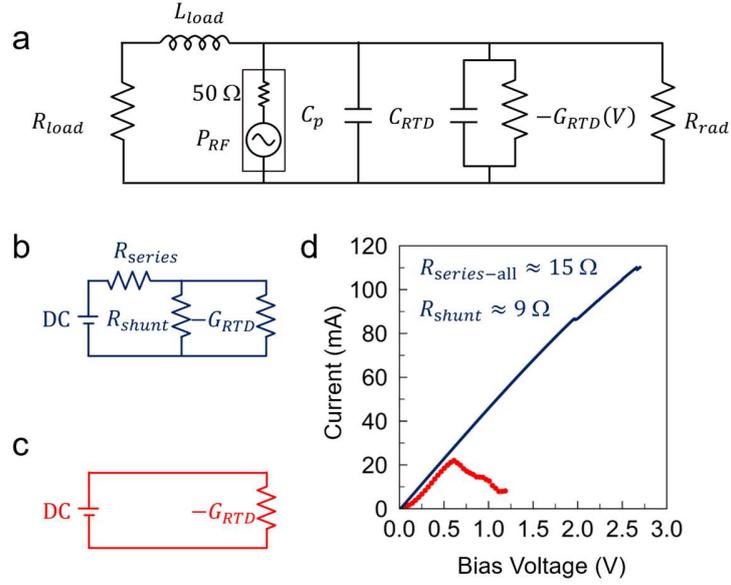

**Extended Data Fig. 5 | Circuit diagram in simulation and *I-V* characteristics. a,** Equivalent circuit diagram of the RTD model used for simulation. RTD is connected in parallel with an RF source having an internal resistance of 50 Ω, resonator load, parasitic capacitance, and radiation resistance. **b,** DC circuit diagram of RTD with shunt resistors and series resistors from the bias tee. **c,** DC circuit diagram of RTD without a shunt resistor and series resistors from the bias tee. **d,** Comparison of measured RTD *I-V* curve with and without shunt resistors and series resistor

| Function | Parameter | Value | Unit |
|---|---|---|---|
| $I_{RTD}(V)$ | $S_{RTD}$ | 2.7 | $\mu m^2$ |
| | $C_1$ | 0.03 | $A \cdot V^{-1} \cdot \mu m^{-2}$ |
| | $C_2$ | 5.5 | $V^{-1}$ |
| | $V_1$ | 0.62 | $V$ |
| | $V_2$ | 0.24 | $V$ |
| | $C_3$ | 0.0015 | $A \cdot V^{-i} \cdot \mu m^{-2}$ |
| | $i$ | 5.4 | – |

**Extended Data Table 1 | Parameters of $I_{RTD}(V)$**

**a** -12 dBm @~0.188 THz

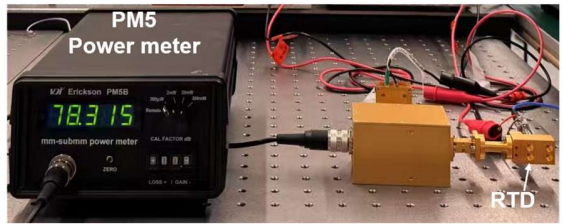

**b** 1.3 dBm @~0.188 THz

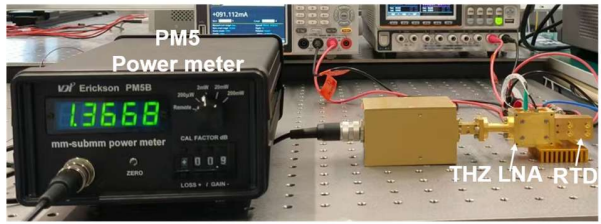

**c** −19.1 dBm @$f_{RF}$ = 20 GHz

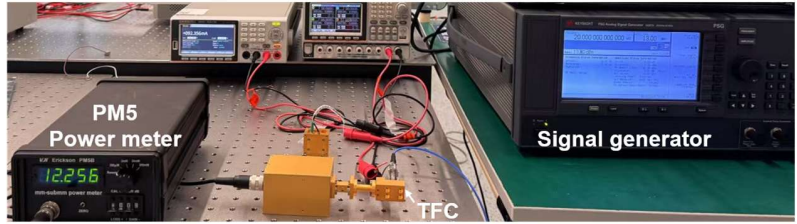

**d** −2.5 dBm @$f_{RF}$ = 20 GHz

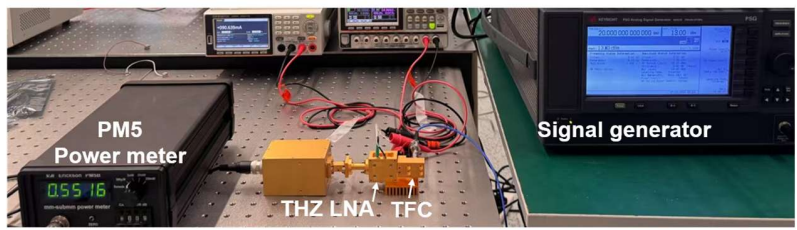

**Extended Data Fig. 6 | Output power measurement. a,** Power measurement of free-running RTD. **b,** Power measurement of free-running RTD with THz LNA. **c,** Power measurement of RTD TFC under 20 GHz 13 dBm injection. **d,** Power measurement of RTD TFC under 20 GHz 13 dBm injection with THz LNA.

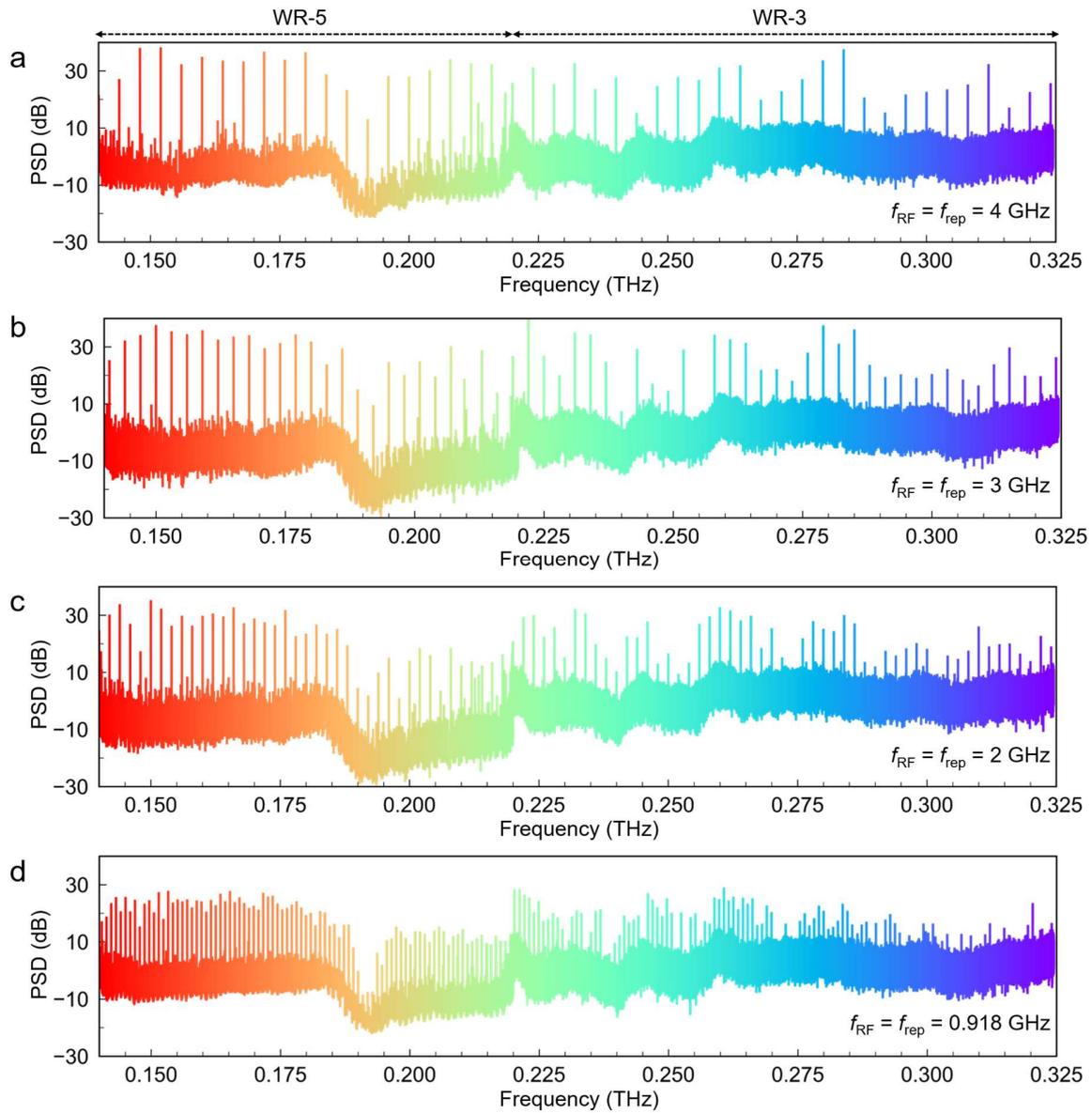

**Extended Data Fig. 7 | Measured spectra of TFC with different RF frequencies and fixed RF injection power of 18 dBm. a,** $f_{RF}$ = 0.918 GHz, **b,** $f_{RF}$ = 2 GHz, **c,** $f_{RF}$ = 3 GHz, and **d,** $f_{RF}$ = 4 GHz. The measurements were performed using two different frequency extenders (WR-5 & WR-3 waveguide types) to cover the spectral ranges of 0.14–0.22 THz and 0.22–0.325 THz. Notably, comb modes are observed at frequencies exceeding 0.3 THz, which is significantly higher than the fundamental frequency of the RTD oscillator. This result highlights the potential for generating TFCs at frequencies far beyond the device's fundamental mode. Furthermore, the ability to generate a stable TFC at different repetition frequencies demonstrates that the comb is widely tunable through the active mode-locking mechanism.

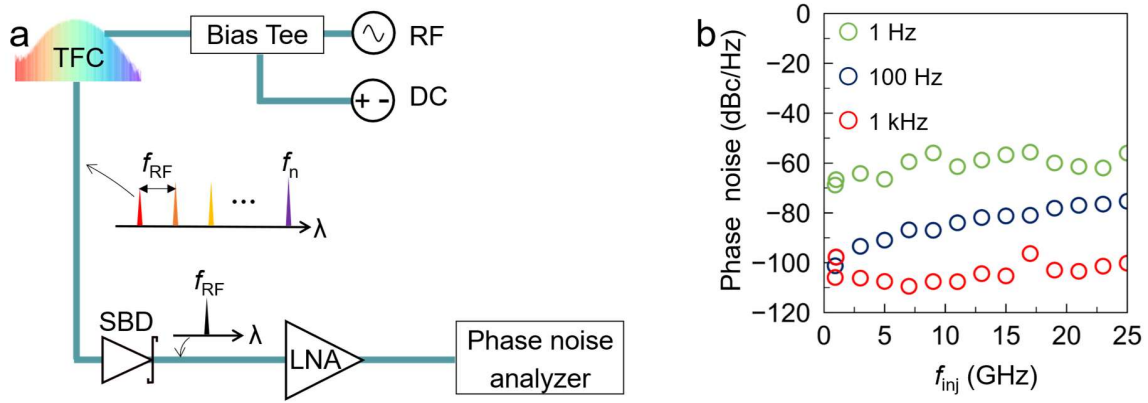

**Extended Data Fig. 8 | Phase noise measurement. a,** Schematic of the TFC Phase noise measurement setup. The beat note, namely $f_{RF}$, is detected at SBD and transmitted to the phase noise analyzer. **b,** Measured phase noise dependence on $f_{RF}$ at 1 Hz, 100 Hz, and 1 kHz, respectively.

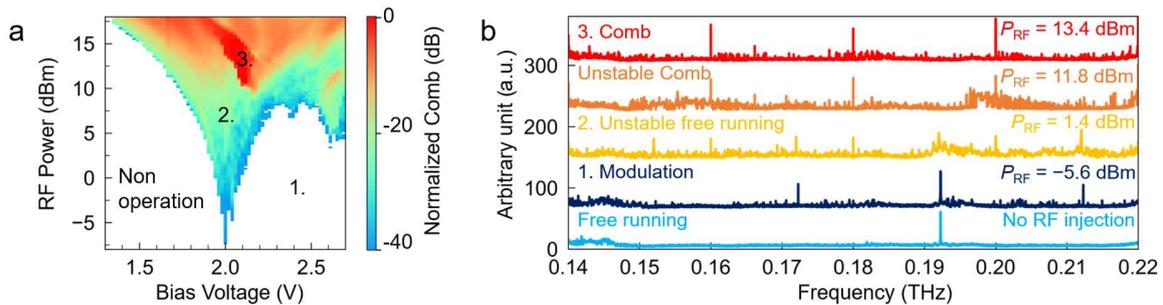

**Extended Data Fig. 9 | Phase diagram of the measured comb power as a function of RF injection power and bias voltage. a,** Total comb power at 0.14, 0.16, 0.18, 0.20, and 0.22 THz when $f_{RF}$ = 20 GHz. **b,** Different operation conditions with different RF power.

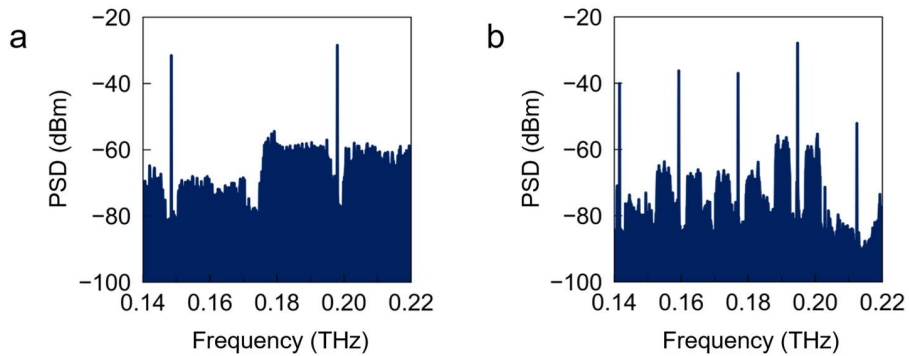

**Extended Data Fig. 10 | Transmitted signal spectrum in wireless communication. a,** Single channel with 32-QAM 24 Gbaud communication when $f_{RF}$ = 49.5 and $f_5$ = 0.198 THz. **b,** Three-channel with 16-QAM 4 Gbaud communication when $f_{RF}$ = 17.7 GHz and $f_9$ = 0.159, $f_{10}$ = 0.177, and $f_{11}$ = 0.195 THz.

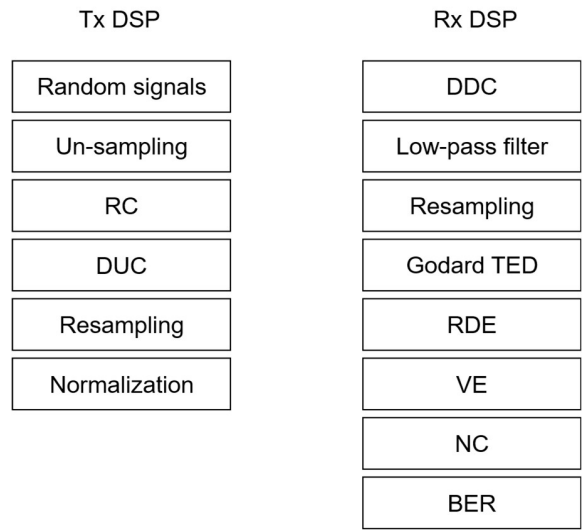

**Extended Data Fig. 11 | DSP flow of Tx and Rx in wireless communication**

| Technology type | Integration | Pump type | Operation temperature | Comb span (THz) | Repetition rate (GHz) | Phase noise @100 Hz (dBc/Hz) | Linewidth (Hz) | Reference |
|---|---|---|---|---|---|---|---|---|
| **RTD** | **Yes** | **RF** | **Room temperature (R.T.)** | **0.14-0.325** | **0.9-49.5** | **-100** | **1** | **This work** |
| RTD | No (Quasi-optical external cavity) | No input | R.T. | 0.293-0.303 | 1 | N/A | 1 | 19 |
| DFG-QCL | Yes | No input | R.T. | 2.2-3.3 | 157 | N/A | N/A | 7 |
| QCL | Yes | RF | 20 K | 2.45-2.55 | 13.3 | N/A | N/A | 5 |
| QCL | Yes | No input | < 55 K | 1.64-3.35 | N/A | N/A | 980 | 23 |
| QCL | Yes | No input | 50 K | 3.25-3.5, 3.65-3.85 | 6.8 | N/A | 1530 | 6 |
| CMOS | Yes | RF | R.T. | 0.22-0.32 | 10 | -50 | N/A | 13 |
| CMOS &PIN-Diode | Yes | Digital pulse | R.T. | 0.03-1.1 | 15 | -50 | 1 | 8 |

**Extended Data Table 2 | Summary of the key metrics of TFC performances.**